# Is the Rayleigh-Sommerfeld diffraction always an exact reference for high speed diffraction algorithms?

SOHEIL MEHRABKHANI,[1,2] AND THOMAS SCHNEIDER[1,2]

[1]Terahertz-Photonics group, Institut für Hochfrequenztechnik, TU Braunschweig, Schleinitzstraße 22, 38106 Braunschweig
[2]Lena, Laboratory for Emerging Nanometrology
*Corresponding author: soheil.mehrabkhani@ihf.tu-bs.de

**Abstract**: In several areas of optics and photonics like wave propagation, digital holography, holographic microscopy, diffraction imaging, biomedical imaging and diffractive optics, the behavior of the electromagnetic waves has to be calculated with the scalar theory of diffraction by computational methods. Many of these high speed diffraction algorithms based on a fast Fourier transformation are in principle approximations of the Rayleigh-Sommerfeld Diffraction (RSD) theory. However, to investigate their numerical accuracy, they should be compared with and verified by RSD. All numerical simulations are in principle based on a sampling of the analogue continuous field. In this article we demonstrate a novel validity condition for the well-sampling in RSD, which makes a systematic treatment of sampling in RSD possible. We show the fundamental restrictions due to this condition and the anomalies caused by its violation. We also demonstrate that the restrictions are completely removed by a sampling below the Abbe resolution limit. Furthermore, we present a very general unified approach for applying the RSD outside its validity domain by the combination of a forward and reverse calculation.

## 1. Introduction

The Rayleigh-Sommerfeld diffraction (RSD) integral is used for the calculation of scalar wave propagation. In contrast to approximations such as Fresnel or Fraunhofer diffraction, the RSD gives an exact solution for the output field of a given input field [1-3]. However, to the best of our knowledge there is no general analytical solution for the calculation of an exact RSD. Therefore, numerical methods have to be used. In these methods, the RSD is treated as a Riemann integral, which has to be discretized. Thus, with usual computational power, only high speed algorithms make the utilization of the diffraction theory possible. These high speed algorithms use approximations of the RSD integral such as a quadratic phase [2, 4] or a frequency-cut in convolution RSD [5-7] and in the angular spectrum method (ASM) [6, 8-11]. Whereupon the latter is only valid for small propagation distances [2, 12]. Contrary to the Kirchhoff solution of the diffraction problem [2, 3, 13], the mathematical solution of the RSD is not inconsistent when the observation point is close to the diffracting screen. Thus, the accuracy of the high-speed alternatives, even for very short propagation distances, could be verified by referencing them with exact solutions given by the RSD. Since there exist only a few analytical solutions, an alternative is to compare them with a discretized RSD. However, here we will demonstrate that the use of the discretized RSD can cause enormous calculation errors. We will give the boundary conditions for the usage of the discretized RSD and show possibilities to completely remove the calculation errors.

## 2. Sampling condition for the Rayleigh-Sommerfeld-diffraction

Figure 1 shows a typical setup for the calculation of the diffraction problem. A coherent source, a laser, illuminates a small object in plane 1. A discretized sensor like a CCD camera would see a diffracted image of the object in plane 2 or 3.

According to the Rayleigh-Sommerfeld diffraction integral the field distribution in an output plane 2 in the distance $z_{12}$, parallel to the input plane is [1-3]:

$$u_2(\vec{R}_2) = -\frac{1}{2\pi}\iint_\Omega d^2r_1 \alpha u_1(\vec{r}_1) e^{ik|\vec{R}_2-\vec{r}_1|} \qquad (1)$$

$$\vec{R}_2 = \vec{r}_2 + z_{12}\hat{e}_z;\ \alpha = \left(ik - \frac{1}{|\vec{R}_2-\vec{r}_1|}\right)\frac{z_{12}}{|\vec{R}_2-\vec{r}_1|^2}$$

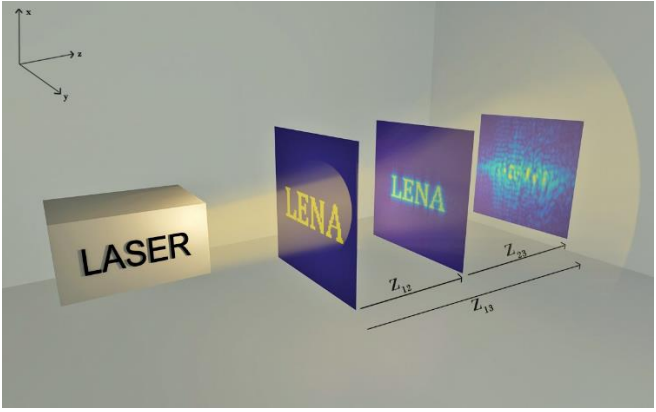

Figure 1: Coherent imaging of an object in the first (object) plane. The second and third planes refer to the diffracted images of the object at the distance $z_{12}$ and $z_{13}$ from the object plane, respectively.

Where $u_1(\vec{r}_1)$ is the field distribution in the input (object) plane 1 and $k$ is the wave number. The two vectors $\vec{r}_1, \vec{r}_2$ are position vectors in plane 1 or 2, respectively.
For a numerical treatment of Eq.1, not only the input and output plane but as well the propagation dependent harmonic term $e^{ik|\vec{R}_2-\vec{r}_1|}$ has to be sampled, according to the well-known Nyquist sampling criterion [14]. Thus, the sampling frequency must at least correspond to twice the highest frequency contained in the harmonic term. If we consider the phase of the propagation term as:

$$\varphi = k|\vec{R}_2 - \vec{r}_1| = k\sqrt{R_2^2 + r_1^2 - 2\vec{R}_2 \cdot \vec{r}_1} \qquad (2)$$
$$= k\sqrt{x_2^2 + y_2^2 + z_{12}^2 + x_1^2 + y_1^2 - 2(x_1 x_2 + y_1 y_2)}$$

with the transversal Cartesian coordinates in the input $(x_1, y_1)$ and output $(x_2, y_2)$ plane, the derivative of the phase $\varphi$ results in the spatial frequency of the propagation phase $f$:

$$f_x = \frac{1}{2\pi}\left|\frac{\partial \varphi}{\partial x_1}\right| = \frac{k|x_1 - x_2|}{2\pi\sqrt{(x_1 - x_2)^2 + (y_1 - y_2)^2 + z_{12}^2}} \qquad (3)$$

This spatial frequency is a monotonically increasing function of $|x_1 - x_2|$. To get its maximum value the conditions $|y_1 - y_2| = 0$ and $|x_1 - x_2| = x_{1fp} + x_{2fp}$ must be fulfilled. Here $x_{1fp}$ and $x_{2fp}$ are the distances of the farthest point from the center in $x$-direction in the relevant computational plane 1 or 2 with nonzero amplitude value (see Fig. 1). Due to zero values of the amplitude around the boundary, the maximum width can be smaller than the real width of the computational domain. Thus, it follows for the maximum spatial frequency $f_{x,max}$ in $x$-direction:

$$f_{x,max} = \frac{1}{2\pi}\left|\frac{\partial \varphi}{\partial x_1}\right|_{max} = \frac{x_{1fp} + x_{2fp}}{\lambda\sqrt{(x_{1fp} + x_{2fp})^2 + z_{12}^2}} \qquad (4)$$

The sampling frequency $f_s$ is related to the sampling spacing via $\delta x_1 = 1/f_s$. Thus, it follows for the sampling according to the Nyquist criterion ($f_s \geq 2f_{max}$):

$$\frac{1}{\delta x_1} \geq 2\frac{x_{1fp} + x_{2fp}}{\lambda\sqrt{(x_{1fp} + x_{2fp})^2 + z_{12}^2}} \qquad (5)$$

Therefore, the validity condition for the numerical treatment of the RSD can be written as:

$$z_{12}^2 \geq \left(\frac{4\delta x_1^2}{\lambda^2} - 1\right)(x_{1fp} + x_{2fp})^2 \qquad (6)$$

Consequently, for a given sampling spacing $\delta x_1, \delta y_1$ in the object plane and a maximum size $x_{1fp}, y_{1fp}$ of the object and its image $x_{2fp}, y_{2fp}$ in the output computational plane, there is a critical minimum propagation distance $z_1$ allowed, in which the output plane can be calculated by:

$$z_{1cx}^2 = \left(\frac{4\delta x_1^2}{\lambda^2} - 1\right)(x_{1fp} + x_{2fp})^2 \qquad (7)$$

An analog derivation results in a similar condition for the $y$-direction:

$$z_{1cy}^2 = \left(\frac{4\delta y_1^2}{\lambda^2} - 1\right)(y_{1fp} + y_{2fp})^2 \qquad (8)$$

Thus, the condition for the minimum propagation distance (critical distance) is:

$$z_{1c} = max(z_{1cx}, z_{1cy}) \qquad (9)$$

The critical distance is the minimum distance in which an output field (image) can be numerically calculated by RSD without violating the Nyquist criterion for the interplay of the sampling conditions of input and output planes and the distance. The reconstruction of the object from the diffracted image is only possible, if the Nyquist theorem is fulfilled in the reverse direction too. This results in equations analogous to (7), (8) and (9) with reversed index 1 and 2.
Thus, the total critical distance $z_c$ for a forward and reverse transformation of the field has to be at least $z_c = max(z_{1c}, z_{2c})$, which is the proposed validity condition for a numerical treatment of the RSD.
According to the validity condition, the correct calculation of the diffraction pattern and also the reconstruction of the original object from the achieved pattern may be performed if the propagation distance is longer than the critical distance:

$$z_{12} > z_c \qquad (10)$$

However, it does not mean that the full information of the object can be obtained from the diffracted image in the output plane. This depends on the numerical aperture of the output plane as well.

## 3. Investigation of the validity condition

### A. Anomalies if $z < z_c$

To show the influence of the violation of the aforementioned validity condition we have used an amplitude object (constant phase) as described in Fig. 2 (a) and (b) and calculated its diffraction pattern at a distance $z < z_c$. For the sake of simplicity but without loss of generality, the phase distributions are compared indirectly. Thus, the magnitude and real part of the complex amplitude will be discussed throughout the paper. As can be seen in Fig. 2, for this object both values are identical because it has a zero phase. The color-bar shows the normalized amplitude and real part values.

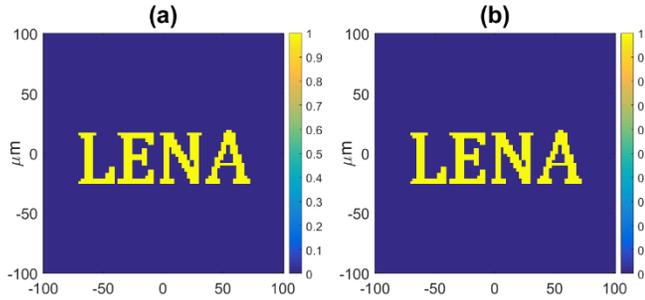

Figure 2: Input object plane, (a) magnitude and real part (b) of the complex amplitude. If not otherwise stated, the following parameters were used for the presented simulations: Wavelength $\lambda = 0.633\ \mu m$, sampling spacing in the input plane $\delta x_1 = \delta y_1 = 0.76\ \mu m$ and in the output plane $\delta x_2 = \delta y_2 = 0.94\ \mu m$. The pixel numbers in the x and y axis as well as in the input and output plane are the same $N_{1,2,x,y} = 265$. The width of the computational domain in the input plane is $P_{x_1} = P_{y_1} = 201\ \mu m$, whereas for the output plane it is $P_{x_2} = P_{y_2} = 250\ \mu m$.

The simulated distance between the object and the image plane is $z_{12} = 20\ \mu m$ which, according to Eqs. (7 - 9), is much smaller than the critical distance $z_c = 534\ \mu m$. The calculated amplitude and real part of the diffracted field $u_{12}$ at a distance $z_{12}$ can be seen in Fig. 3 (a) and (b) respectively. The error due to the violation of the validity condition can be seen by a reconstruction of the original object from the diffracted image $u_{12}$. Thus, the reverse RSD $z_{21}$ has to be applied. Here the term "reverse" instead of "inverse" transform will be used in order to avoid confusion.

The result can be seen in Fig. 3 (c) and (d). A similar pattern like in the original object occurs but, with very strong anomalies. Especially the nonzero values of the field in areas, which originally exhibit zero values, lead to completely wrong results. The correlation coefficient $r$ between $u_1$ and $u_{121}$, i.e. the field after transforming from position 1 to 2 and back to 1 is $r = 0.72$ and $r = 0.75$ for the magnitude of the complex field and its real part, respectively. The obviously strong deviation to the input is a consequence of the insufficiency of the propagation distance $z < z_c$.

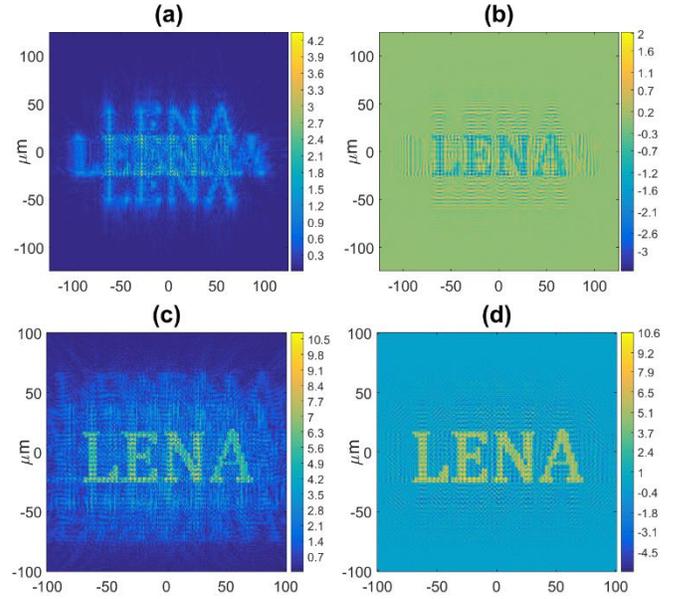

Figure 3: Reconstruction errors for $z < z_c$ (a) magnitude and (b) real part of the complex amplitude of the diffracted image in plane 2. (c) and (d) magnitude and real part of the reconstructed object in plane 1.

### B. Propagation distance $z > z_c$

As derived in section 1, if the condition $z_{12} > z_c$ is satisfied for the numerical calculation of the RSD, there will be no loss of information due to the transformation. For an object with a fixed sampling spacing the critical propagation distance is fixed. Accordingly, if the propagation distance is longer than $z_c$, a correct treatment of the computational RSD should be achieved for the given sampling spacing.

In Fig. 4 (a) and (b) the diffracted image is numerically calculated by the RSD under the assumption that the propagation distance $z_{13} = 730\ \mu m$ satisfies the validity condition. To confirm the correctness of the field $u_{13}$ in the plane 3 as a necessary condition, the reverse transform is considered to reconstruct the input object in the input plane, as reported in Fig. 4 (c) and (d). The correlation coefficient is $r = 0.97$ for both, magnitude and real part of the complex amplitude. Since a small part of the whole information will be lost by the spatial limitation of the computational plane, it is smaller than one. Comparing the correlation coefficients for the reconstructed object in Fig. 3 and Fig. 4 shows a more than 20% improvement. Therefore, at the distance $z_{13}$ almost the whole information of the object plane is preserved. However, in some cases the RSD might be applied as a reference for different algorithms at a propagation distance outside of the validity domain.

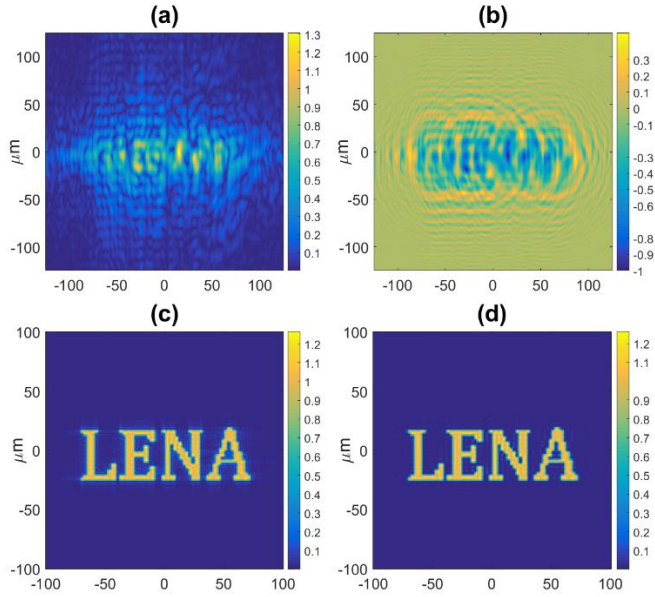

Figure 4: (a), (b) magnitude and real part of the image complex amplitude, respectively for $z_{13} > z_c$. (c) and (d) corresponding reconstructed amplitude and real part of the reconstructed object in the input plane.

Additionally, a good object reconstruction by a forward and reverse calculation is just a necessary, but not a sufficient condition for testing the validity of a diffraction algorithm. It does not necessarily mean, that the output corresponds to the expected result of the diffraction theory. In other words, a combination of an arbitrary propagation operator (physically or non-physically) with its inverse always results in an identity operator, and consequently the reconstruction of the input is expected automatically. Thus in the next subsection a sampling condition, which always satisfies the validity condition and which can be used as a reference for the diffraction will be presented and in section 4 a general procedure, which makes the RSD a feasible method for arbitrary propagation distances will be shown.

**C. Sampling spacing below the Abbe resolution limit for fine structures larger than the Abbe limit**

According to inequality 6, the left side and the second term on the right side are always positive whereas, the first term on the right side can change its sign. For a sampling lower than half of the wavelength $\delta x_1 < \frac{\lambda}{2}$, it will become negative and consequently the inequality will be fulfilled for all propagation distances $z$. Thus, the validity condition is always satisfied, if the sampling spacing of the harmonic term is smaller than the Abbe resolution limit. It should be emphasized that this condition only holds for the harmonic term. As will be shown in subsection D, this does not contradict the Abbe resolution limit.

According to the discussion above, substructures smaller than the Abbe limit could be resolved and consequently be reconstructed by the Nyquist criterion. However, there are different meanings of „reconstruction" in respect to diffraction (restricted due to the Abbe limit) and in respect to the sampling theory, restricted by the Nyquist criterion. In the context of the sampling theory, a direct reconstruction of the original field after sampling will be possible, whereas for the diffraction theory the reconstruction of the original field is indirect, since it takes place after a propagation over the distance $z$. Thus, the sampled data is exposed to the diffraction effect and consequently restricted by the Abbe limit. Eventually, a sampling below the wavelength does not lead to the breaking of the Abbe rule for our approach, but it enables the calculation of a diffracted image without violating the RSD validity condition.

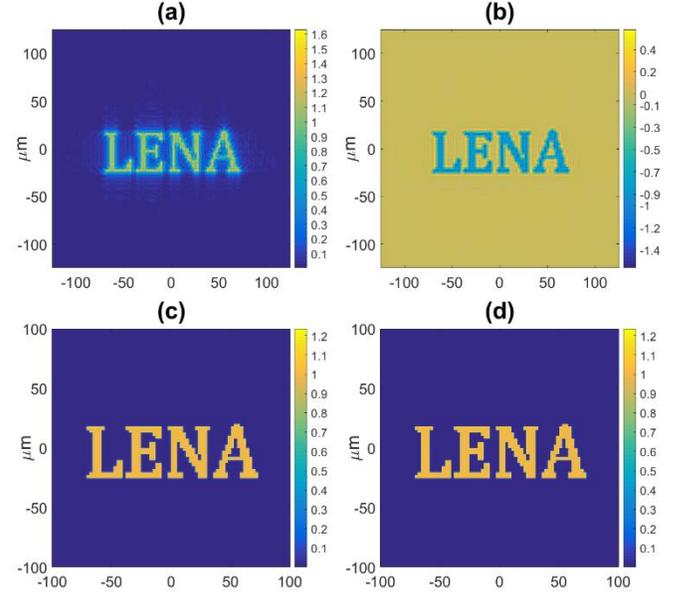

Figure 5: (a) and (b) magnitude and real part of the complex amplitude according to the RSD at the same propagation distance like Fig. 3 but with a sampling spacing of the object and image smaller than the Abbe limit., (c) and (d) reconstruction of the object. The structures in the object are larger than the Abbe limit.

In Fig. 5 (a) and (b) the diffracted image for the same simulation parameters like in Fig. 3 can be seen ($z_{12} < z_c$), except that for this simulation the object was sampled with a sampling spacing below the Abbe limit. However, the structures in the object are still larger than the Abbe limit. The reconstructed object is shown in Fig. 5 (c) and (d). The correlation coefficient between the reconstructed and input object is $r = 0.997$ for both the magnitude and the real part of the complex amplitude. The minor loss of information is just due to the limited aperture. Thus, the sampling of the harmonic term with a sampling spacing smaller than the Abbe limit can be used as a reference for the evaluation of the quality of the numerical calculations.

**D. Sampling spacing below the Abbe resolution limit for fine structures smaller than the Abbe limit**

To investigate the effect of the sampling below the Abbe limit for under Abbe limit structures, the input area, the output area and the sampling spacing have been rescaled (10 and 20 times smaller than the object in Fig. 2), so that both the object's fine structures and the sampling spacing are below the Abbe limit. In Fig. 6 (a, b) and (c, d) the rescaled object and the reconstructed object are shown respectively.

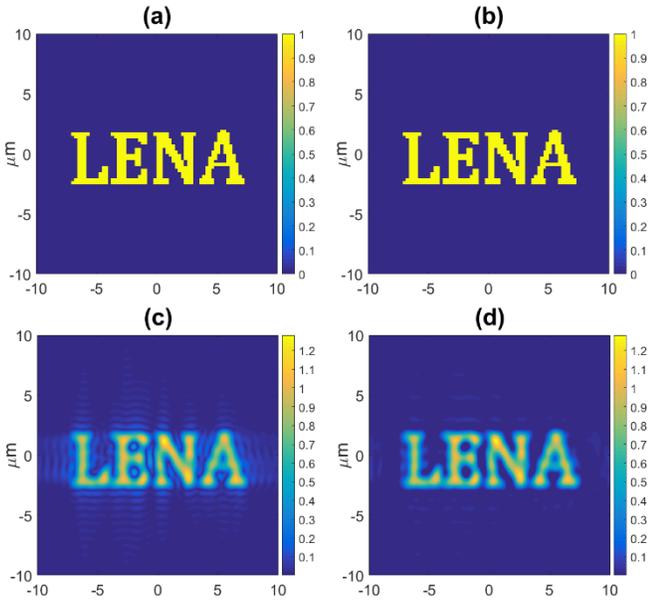

Figure 6: (a), (b) magnitude and the real part of the complex amplitude for the rescaled input object. (c) and (d) magnitude and real part of the complex amplitude for the reconstructed object for a sampling spacing below the Abbe limit $\delta x_1 = \delta y_1 = 0.025$ μm $< \frac{\lambda}{2} = 0.32$ μm.

As can be seen, due to the violation of the Abbe resolution limit, the fine structures of the object in Fig. 6 cannot be resolved anymore. The calculated correlation coefficients are $r = 0.88$ and $r = 0.89$ for the magnitude and the real part of the complex amplitude respectively.

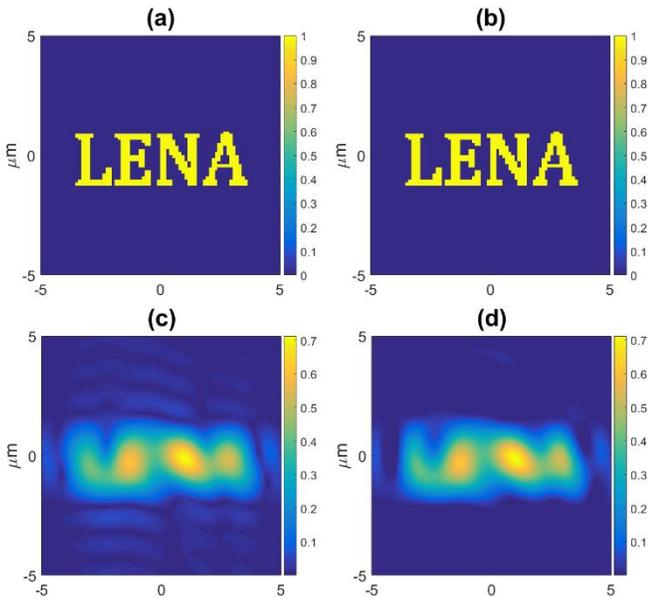

Figure 7: (a) and (b) magnitude and real part of the complex amplitude of the input object. (c) and (d) magnitude and real part of the complex amplitude for the corresponding output by a sampling below the Abbe limit $\delta x_1 = \delta y_1 = 0.013$ $\mu m << \frac{\lambda}{2} = 0.32$ $\mu m$.

For a further reduction of the fine structures in the object the effect is increased as can be seen in Fig. 7 (a-d). The calculated correlation coefficients are only $r = 0.62$ and $r = 0.63$. Thus, a subwavelength sampling in RSD cannot retain the full object information, if the fine substructures are below the Abbe limit.

## 4. General solution for removing the limitation of the propagation distance

In practical applications the sampling of the object is restricted by a minimum spacing as a consequence of the limited pixel size of a given CCD camera. Here a general solution for an arbitrary distance from the object will be presented.

The RSD is a linear operator $\mathcal{R}$, which transforms an input field $u_1(\vec{r})$ over a propagation distance $z_{12}$ into the field $u_{12} = \mathcal{R}_{12}\{u_1\}$. Theoretically, for an unlimited aperture the information in the input plane $u_1$ is completely conserved in the diffracted image $u_{12}$. Therefore, the field $u_1$ can be reconstructed from the field $u_{12}$ by a reverse application of the RSD with $z \to -z$. The combination of the forward operator $\mathcal{R}_{12}$ and the reverse operator $\mathcal{R}_{21}$ ($u_1 = \mathcal{R}_{21}\{u_{12}\}$) of the field is an identical operator $\mathcal{R}_{21}\mathcal{R}_{12} = \mathbb{1}$. Thus, it can be written that $\mathcal{R}_{21}\mathcal{R}_{12}\{u_1\} = \mathbb{1}\{u_1\} = u_1$ (for $z_{12} > z_{12_c}$).

If the validity condition is not satisfied $z_{12} < z_{12_c}$, a complete reconstruction is not possible and $\mathcal{R}_{21}\mathcal{R}_{12} \neq \mathbb{1}$. If a set $\Upsilon$ of all propagation distances satisfying the validity condition is introduced, it follows that the reverse transform is not an inverse transform if $z \notin \Upsilon$. Although analytically the reverse and the inverse transforms are identical. Thus, a perfect reconstruction of all the information in the object is only possible if $\mathcal{R}_{21}\mathcal{R}_{12} = \mathbb{1}$. Therefore an RSD operator, which satisfies $\mathcal{R}_{21}\mathcal{R}_{12} = \mathbb{1}$ for $z \notin \Upsilon$ has to be found.

As described in the last section, at the distance $z_{13} \in \Upsilon$, the reconstruction of the object is almost perfect but, outside the validity condition $z_{12} \notin \Upsilon$, the whole object information cannot be retrieved from the field $u_{12}$. Thus for a general solution, the following approach is proposed: in a first step the image at a longer distance which satisfies the validity condition and identity relation $z_{13} \in \Upsilon$, is calculated with the additional property $z_{13} - z_{12} \in \Upsilon$. In a second step a new propagation distance $z_{23} = z_{13} - z_{12} \in \Upsilon$ will be calculated with $\mathcal{R}_{23} = \mathcal{R}_{32}^{-1}$. The operator $\mathcal{R}_{32}$ transforms the field $u_{13}$ at $z_{13}$ to the field $u_{132}$ at a shorter distance $z_{12}$. Thus, the operator $\mathcal{R}_{132} = \mathcal{R}_{32}\mathcal{R}_{13}$, which transforms the field $u_1$ to the field $u_{132} = \mathcal{R}_{132}\{u_1\}$ at the distance $z_{12}$ is introduced. Although $z_{12} \notin \Upsilon$, it can be easily shown that $\mathcal{R}_{231}$ satisfies the identity relation as follows:

The reverse of the operator $\mathcal{R}_{132}$ is the operator $\mathcal{R}_{231}$. According to operator theory [16]:

$$\mathcal{R}_{231}\mathcal{R}_{132} = (\mathcal{R}_{31}\mathcal{R}_{23})(\mathcal{R}_{32}\mathcal{R}_{13}) = \mathcal{R}_{31}\mathcal{R}_{23}\mathcal{R}_{32}\mathcal{R}_{13}$$

$$= \mathcal{R}_{13}^{-1}\mathcal{R}_{32}^{-1}\mathcal{R}_{32}\mathcal{R}_{13} = \mathcal{R}_{13}^{-1}\mathbb{1}\mathcal{R}_{13} = \mathbb{1} \quad (11)$$

Which means the reverse and inverse transforms are the same $\mathcal{R}_{231} = \mathcal{R}_{132}^{-1}$. If the loss of information due to the limited aperture for practical applications is neglected, the new image $u_{132}$ contains all information from $u_1$.

The operator $\mathcal{R}_{132}$ depends on two propagation variables $z_{12}$ and $z_{13}$. The first is the real variable, which determines the distance between the object and the image. The second is just an arbitrary parameter which has to fulfill the condition $z_{13} \in$

$\Upsilon$, $z_{13} - z_{12} \in \Upsilon$. Thus, the set $\Upsilon$ has an infinite number of elements, which are all valid. However, a cutting of diffracted field values due to the limited size of the computational plane 3 leads to a loss of information. Thus, for a fixed value of the pixel size and pixel number, the optimal choice for the propagation distance $z_{13}$ is the minimum allowed value. In Fig. 8 the calculated field $u_{132}$ at the distance $z_{12} \notin \Upsilon$ is compared with the field $u_{12}$ at the same distance. This field $u_{12}$ was calculated for a sampling spacing below the Abbe limit and can be used as a reference, as discussed in section 3C.

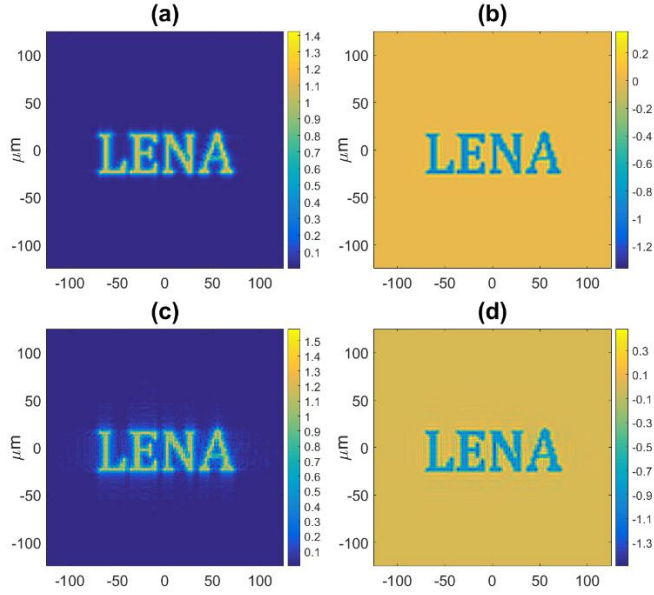

Figure 8: (a) and (b) magnitude and real part of the complex amplitude $u_{132}$, (c) and (d) magnitude and real part of the complex amplitude $u_{12}$ by sampling below the Abbe limit, used as a reference.

The correlation coefficient for the magnitude and real part of the amplitude are $r = 0.97$ respectively $r = 0.95$, which shows a remarkable improvement compared to $r = 0.65$ and $r = 0.49$ for the case of applying the conventional RSD $\mathcal{R}_{12}$. If we compare the image in Fig. 3 (a), (b) with the below Abbe sampling image in Fig. 8 (c) and (d), we have a 32% improvement in the magnitude and 46% in the real part of the amplitude.

In Fig. 9 the reconstruction of the object $u_{13231}$ by the use of the operator $\mathcal{R}_{132}$ for the forward and $\mathcal{R}_{231}$ for the backward propagation is presented. The correlation coefficients for the magnitude and the real part of the complex amplitude are $r = 0.97$. Again the validity and capability of the proposed approach can be seen.

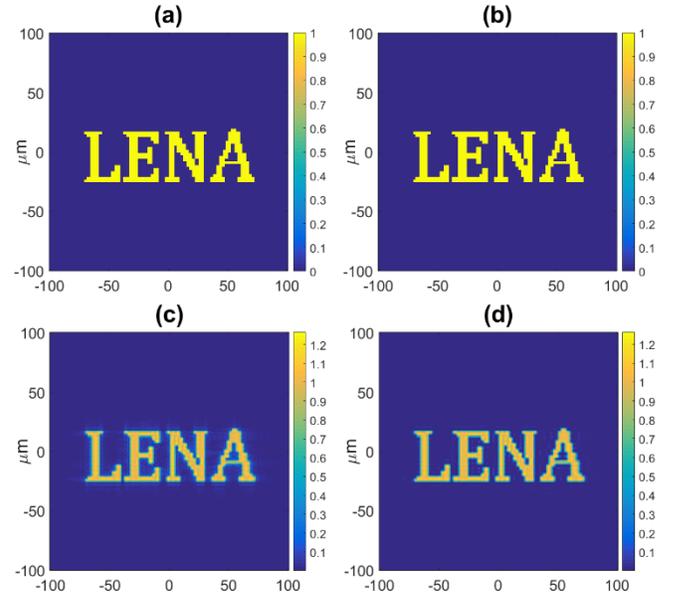

Figure 9: (a), (b) amplitude and real part of the input object (c), (d) magnitude and real part of the amplitude for the reconstructed object.

## 5. Conclusion

In this paper the numerical treatment of the Rayleigh-Sommerfeld diffraction was investigated in detail. A validity condition for the numerical calculation was derived. As have been shown, for a fixed sampling spacing in the computational domain, the allowed propagation distance is restricted to a minimum value. However, the restriction can be completely removed if the sampling spacing (not the structure in the object) is lower than the Abbe limit. As have been shown, this results in the maximum obtainable information in the output plane under the consideration of the limited computational domain, and was therefore used as a reference. Moreover, a very general approach for the calculation of the output field for arbitrary propagation distances was presented. This operator is based on a combination of forward and reverse RSD transforms and leads to very high correlation coefficients of $r = 0.97$. An about 30% improvement in the magnitude of the amplitude and about 45% for the real part confirms the reliability of the new operator. A comparison of the results of the below Abbe limit sampling with the results of the composed operator is an additional verification of both methods and the consistency of the theoretically derived validity condition for the RSD. The developed approach can be used as a reference for the testing of high speed algorithms and other methods, which are based on the approximation of the exact scalar diffraction theory.

**Acknowledgment**. We acknowledge the financial support by the Ministry of Science and Culture of Lower Saxony in the framework of QUANOMET. We like to thank Ali Dorostkar for very fruitful discussions, Taranom Akbari for the creation of Fig. 1 and Julia Böke, Okan Özdemir and Stefan Preußler for their support during the writing of the paper.